\documentclass[a4paper]{jpconf}
\usepackage{graphicx}
\usepackage{epsf}
\usepackage{axodraw}
\usepackage{cite}

\begin{document}
\title{SUSY CP phases and asymmetries at colliders }

\author{Olaf Kittel}

\address{
Departamento de F\'isica Te\'orica y del Cosmos and CAFPE, \\
Universidad de Granada, E-18071 Granada, Spain}


\begin{abstract}
In the Minimal Supersymmetric Standard Model,
physical phases of complex parameters lead to CP violation.
We show how triple products of particle momenta or spins can be
used to construct asymmetries, that allow us to probe these CP phases.
To give specific examples, we discuss the production of neutralinos at the
International Linear Collider~(ILC).
For the Large Hadron Collider~(LHC), 
we discuss CP asymmetries in squark decays, and 
in the tri-lepton signal.
We find that the CP asymmetries can be as large as 60\%.

\end{abstract}

\section{Introduction}
\smallskip

Supersymmetry~(SUSY)~\cite{mssm} 
is a very well motivated theory to extend the Standard
Model~(SM) of particle physics~\cite{Buchmuller:2006zu}.
SUSY models are not only favored by gauge coupling unification and naturalness 
considerations, but are also attractive from the cosmological point of view.
For example, the lightest SUSY particle~(LSP)  
is a good dark matter candidate, if it is stable, massive and weakly
interacting~\cite{Goldberg:1983nd,RelicCP}. 
Most interestingly, SUSY models provide a number of new parameters,
among some having physical phases which cause manifest 
CP violation~\cite{Haber:1997if}.
Remarkably in the SM, the single CP phase in the quark mixing matrix,
which is currently confirmed in $B$ meson 
experiments~\cite{Amsler:2008zz,Buras:2005xt},
cannot explain the observed baryon asymmetry of the 
universe~\cite{Csikor:1998eu}.
Additional sources of CP violation in models beyond the SM are 
required~\cite{Riotto:1998bt}.

\smallskip

In the MSSM, a set of remaining complex parameters is obtained
after absorbing unphysical phases of parameters by redefining particle fields.
In the literature, the complex parameters  are usually chosen to be the 
higgsino mass parameter  $\mu$, 
the  ${\rm U(1)}$ and ${\rm SU(3)}$  gaugino mass parameters $M_1$ and $M_3$, 
respectively, and the trilinear scalar coupling parameters $A_f$
\begin{eqnarray}
\label{eq:phases}
\mu = |\mu| e^{i \varphi_\mu}, \quad  
M_1 = |M_1| e^{i \varphi_{M_1}}, \quad
M_3 = |M_3| e^{i \varphi_{M_3}},\quad
A_f = |A_f| e^{i \varphi_{A_f} }.
\end{eqnarray}
The SUSY CP phases of these parameters lead to theoretical predictions for the 
electric dipole moments~(EDMs) of electron, neutron and that of the 
atoms $^{199}$Hg and $^{205}$Tl,
which can be  (sometimes orders of magnitude) beyond the current experimental upper 
bounds~\cite{Amsler:2008zz,Ellis:2008zy,Choi:2004rf}.
These strong bounds suggest that either the SUSY CP phases are
severely suppressed (in particular $\varphi_\mu <0.1\pi$), 
the SUSY particles are very heavy (e.g. the first two generations of sfermions with 
$m_{\tilde f}>10$~TeV), or different loop-contributions to the EDMs cancel
accidentally.  
However, these solutions  require a fine-tuning of the SUSY parameters, 
and would be unnatural. 
The need of tuning the SUSY parameters and phases to fulfill the
EDM constraints is  referred to as the \emph{SUSY CP problem} in the 
literature~\cite{Abel:2005er,Ellis:1982tk}. 
In order to analyze the problem, it is necessary to independently
measure the SUSY phases at colliders. In particular CP-odd observables are 
needed to find or exclude direct evidence of CP violation.

\smallskip

In this talk we will concentrate on a particular class of 
CP-odd\footnote{Other classes of CP-odd observables would 
  be rate asymmetries
  of cross sections, branching ratios and 
  distributions~\cite{otherCPodd}.
  In addition it should be noted that SUSY phases have large impact on 
  the neutral MSSM Higgs sector. For recent reviews and references, see for 
  example Ref.~\cite{CPreviews}
} 
(T-odd) observables, which can be defined with the help of
triple products~\cite{tripleprods}. Triple products can lead to large
CP asymmetries, since they already appear at tree level
due to spin correlations~\cite{Haber:1994pe}. 
We will only give a few selected examples to motivate the use of triple products,
and cannot give a thorough review of the vast amount of literature in the
field. Therefore we will discuss the production of 
neutralinos~\cite{Bartl:2003tr,Bartl:2003gr,Choi:2003pq} at the
International Linear Collider~(ILC)~\cite{ILC}. 
For the Large Hadron Collider~(LHC)~\cite{LHC}, 
CP asymmetries in top squark decays~\cite{Bartl:2004jr,stop,Ellis:2008hq} 
and in the tri-lepton signal~\cite{trilepton,plan} are discussed.

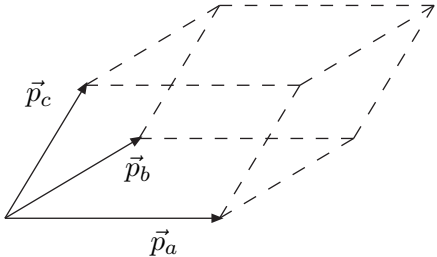
\begin{figure}[t]
\begin{minipage}{12pc}
        \begin{picture}(160,50)(-28,-30)
\scalebox{1}{
\Text(8,42)[lb]{$ \vec p_{c}$}
\Text(65,-4)[rt]{$ \vec p_{a}$}
\Text(45,14)[lb]{$ \vec p_{b}$}
     \LongArrow(0,0)(80,0)
     \LongArrow(0,0)(50,30)
     \LongArrow(0,0)(30,50)
     \DashLine(80,0)(130,30){5}
     \DashLine(80,0)(110,50){5}
     \DashLine(50,30)(130,30){5}
     \DashLine(50,30)(80,80){5}
     \DashLine(30,50)(80,80){5}
     \DashLine(30,50)(110,50){5}
     \DashLine(80,80)(160,80){5}
     \DashLine(130,30)(160,80){5}
     \DashLine(110,50)(160,80){5}
}
                                \end{picture}
\end{minipage}\hspace{6.5pc}
\begin{minipage}[b]{18.4pc}
\caption{\label{Fig1}
  Triple product ${\mathcal T}=(\vec p_{a} \times \vec p_b) \cdot \vec p_c$ of
  three particle momenta (or spin) vectors. The absolute value of the triple product is
  a measure for the volume spanned by the vectors. The sign of the triple product 
  is a measure for their orientation: a negative (positive) sign corresponds to a
  left- (right-) handed system.
}
\end{minipage}
\end{figure}

\section{Triple products and their asymmetries}
\smallskip

Triple products are built up from particle spin or momenta three-vectors, 
\begin{equation}
{\mathcal T} = (\vec p_{a} \times \vec p_b) \cdot \vec p_c,
\label{eq:tripleprod}
\end{equation}
see a schematic picture in Fig.~\ref{Fig1}. Since each of the momentum (spin)
vector changes its sign under a naive time transformation, $t \to -t$, the 
triple product is T-odd.
Thus T-odd asymmetries of the cross section $\sigma$ can be defined
\begin{eqnarray}
 {\mathcal A}_{\rm T} = \frac{\sigma({\mathcal T}>0)
                             -\sigma({\mathcal T}<0)}
                             {\sigma({\mathcal T}>0)
                             +\sigma({\mathcal T}<0)}
      =
          \frac{\int {\rm Sign}[{\mathcal T}]
                 |T|^2 d{\rm Lips}}
           {\int |T|^2 d{\rm Lips}},
\label{eq:Tasymmetry}
\end{eqnarray}
with the amplitude squared $|T|^2$, and the Lorentz invariant phase-space element
$d{\rm Lips}$, such  that $\int |T|^2 d{\rm Lips}=\sigma$. 
The triple product asymmetry is thus an angular distribution
\begin{eqnarray}
 {\mathcal A}_{\rm T} = \frac{N_+ - N_-}{N_+ + N_-},
\label{eq:Tasymmetry2}
\end{eqnarray}
with the number of events $N_+$ ($N_-$) of particle $\vec p_c$
above (below) the plane spanned by $\vec p_{a} \times \vec p_b$.
%
%
The T-odd asymmetry ${\mathcal A}_{\rm T}$ would also be CP-odd, 
if absorptive phases (e.g. from higher order final-state interactions or 
finite-widths effects) can be neglected.
Since the absorptive phases do not change sign under charge conjugation,
they can be eliminated in some cases by defining a genuine CP asymmetry 
\begin{eqnarray}
 {\mathcal A}_{\rm CP} = \frac{1}{2}({\mathcal A}_{\rm T} -\bar{\mathcal A}_{\rm T}),
\label{eq:CPasymmetry}
\end{eqnarray}
of the corresponding  asymmetry $\bar{\mathcal A}_{\rm T}$ for the 
charge conjugated process.

\begin{figure}[t]
\begin{minipage}{16pc}
\includegraphics[height=16pc, width=16pc]{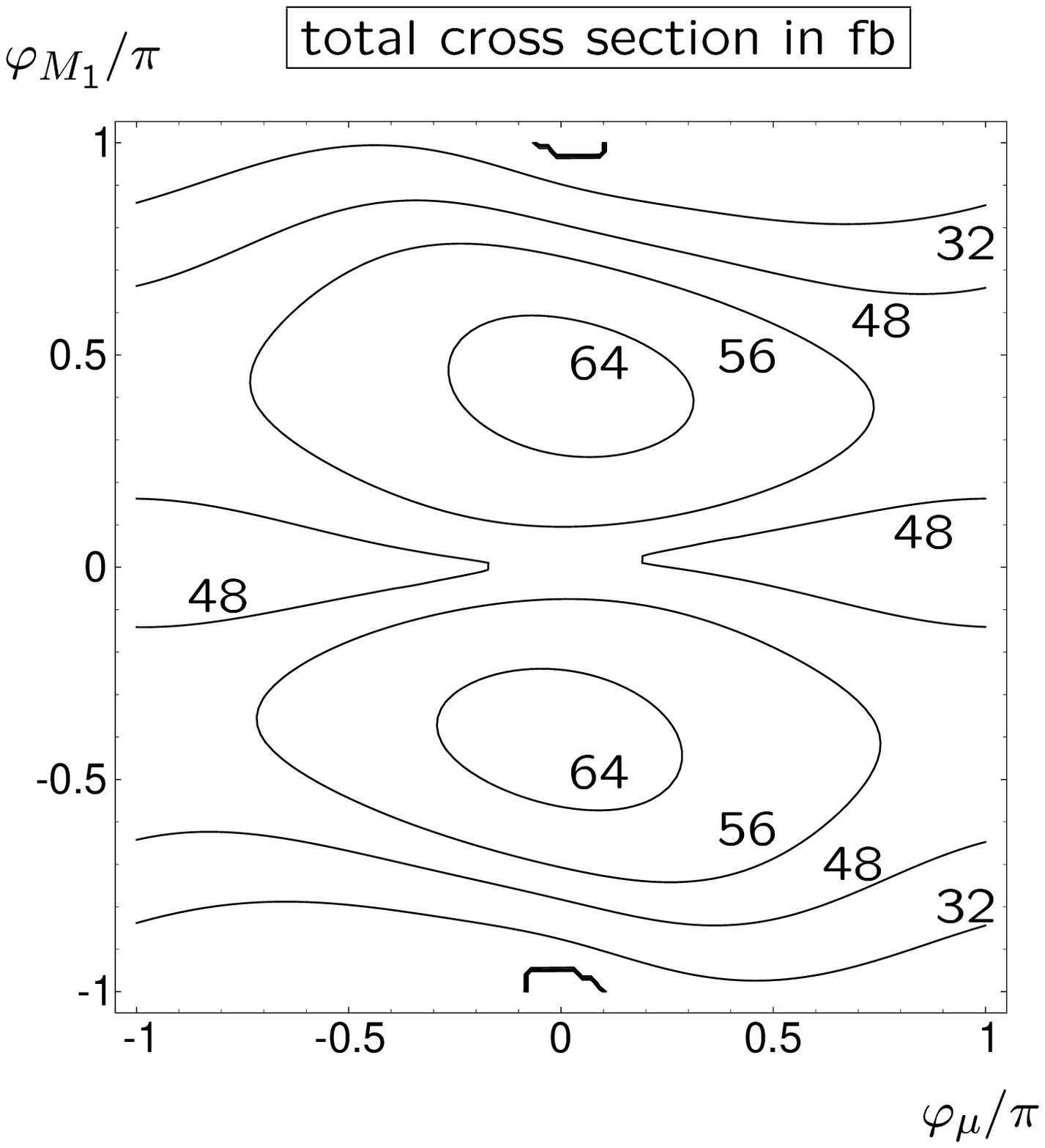}
\put(-180,0){(a)}
\end{minipage}\hspace{4pc}%
\begin{minipage}{16pc}
\includegraphics[height=16pc,width=16pc]{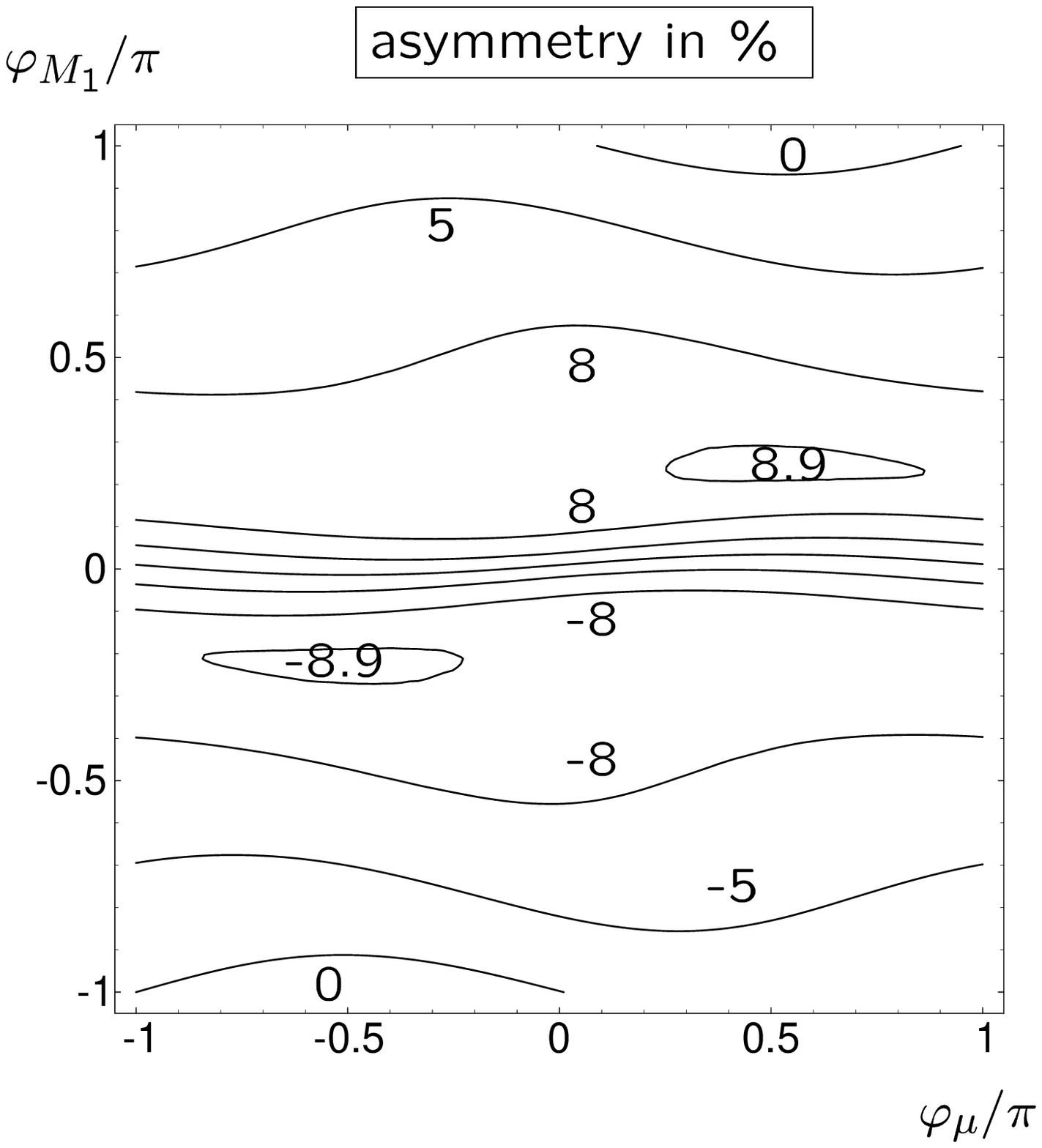}
\put(-180,0){(b)}
\end{minipage} 
\caption{\label{FigNeut} 
        Contour lines of (a) the cross section
        $\sigma_P(e^+e^-\to\tilde\chi^0_1\tilde\chi^0_2)\times
        {\rm BR}(\tilde \chi^0_2\to\tilde\ell_R\ell_1)\times
        {\rm BR}(\tilde\ell_R\to\tilde\chi^0_1\ell_2)$
        with ${\rm BR}(\tilde\ell_R \to\tilde\chi^0_1\ell_2)=1$,
        and (b) the T-odd asymmetry ${\mathcal A}_{\rm T}$,
        in the $\varphi_{\mu}$--$\varphi_{M_1}$ plane, 
        for  $M_2=400$ GeV, $|\mu|=240$ GeV, $\tan \beta=10$, $m_0=100$ GeV,
        $A_{\tau}=-250$ GeV, 
        at  $\sqrt{s}=500$ GeV, and polarized beams
        $(P_{e^-},P_{e^+})=(0.8,-0.6)$~\cite{Bartl:2003tr}.
        }
\end{figure}

\section{Numerical examples for the ILC}
\smallskip

To start with an intuitive example of the use of triple products,
we consider the production of the lightest neutralino pair at the ILC
\begin{eqnarray} 
        e^+ +e^-&\to&\tilde\chi^0_1 + \tilde\chi^0_2.
\label{NEUTproduction}
\end{eqnarray}
Since, due to momentum conservation, the production takes place in a plane,
an additional vector-degree of freedom perpendicular to this plane is needed to built
up a triple product. Indeed, the polarization of each  neutralino normal
to the production plane is CP-sensitive, and vanishes if  CP is conserved,
i.e., for $\phi_{M_1} = \phi_{\mu} = 0$ (modulo $\pi$).
In the following, we analyze the polarization of neutralino $ \tilde\chi^0_2$ 
by its leptonic two-body decay chain~\cite{Bartl:2003tr,Bartl:2003gr,Choi:2003pq}
\begin{eqnarray}
\tilde\chi^0_2 \to \tilde\ell^\pm + \ell^\mp; \qquad 
 \tilde\ell^\pm \to \tilde\chi^0_1 + \ell^\pm; \qquad 
 \ell= e,\mu.
\label{NEUTdecay}
\end{eqnarray}
Now a triple product 
${\mathcal T} = 
(\vec p_{\ell^+} \times \vec p_{\ell^-}) \cdot \vec p_{e^-}$,
see Eq.~(\ref{eq:tripleprod}),
can be formed from the beam momentum $\vec p_{e^-}$
and the two outgoing lepton momenta $\vec p_{\ell^\pm}$.

\medskip

In Fig.~\ref{FigNeut}, we show the phase dependence of the cross section
and the corresponding asymmetry ${\mathcal A}_{\rm T}$~(\ref{eq:Tasymmetry}).  
It is remarkable that the maximal values ${\mathcal A}_{\rm T}\approx\pm9\%$
are not necessarily obtained for maximal CP phases. 
The reason is that the asymmetry ${\mathcal A}_{\rm T}$ is 
proportional to the neutralino spin correlations, which are a product of
a CP-odd term from the production, and a CP-even term from the decay.
Since the  CP-odd (CP-even) factor has
as sine-like (cosine-like) dependence on the phases,
the maxima of ${\mathcal A}_{\rm T}$ are shifted  
towards $\varphi_{M_1}=0$ in Fig.~\ref{FigNeut}(b).
It is interesting to note that the asymmetry can be sizable 
for small values of the phases, which is suggested by the 
EDM constraints. Note also that  the variation of the cross section, 
Fig.~\ref{FigNeut}(a), is more than $100\%$. 
In addition to the CP-sensitive asymmetry, the cross section may 
serve to constrain the phases. Note that the choice of longitudinal
beam polarization $(P_{e^-},P_{e^+})=(0.8,-0.6)$ almost doubles the size 
of the asymmetry and the cross section.

\medskip

\begin{figure}[t]
\begin{minipage}{16pc}
\includegraphics[height=16pc, width=16pc]{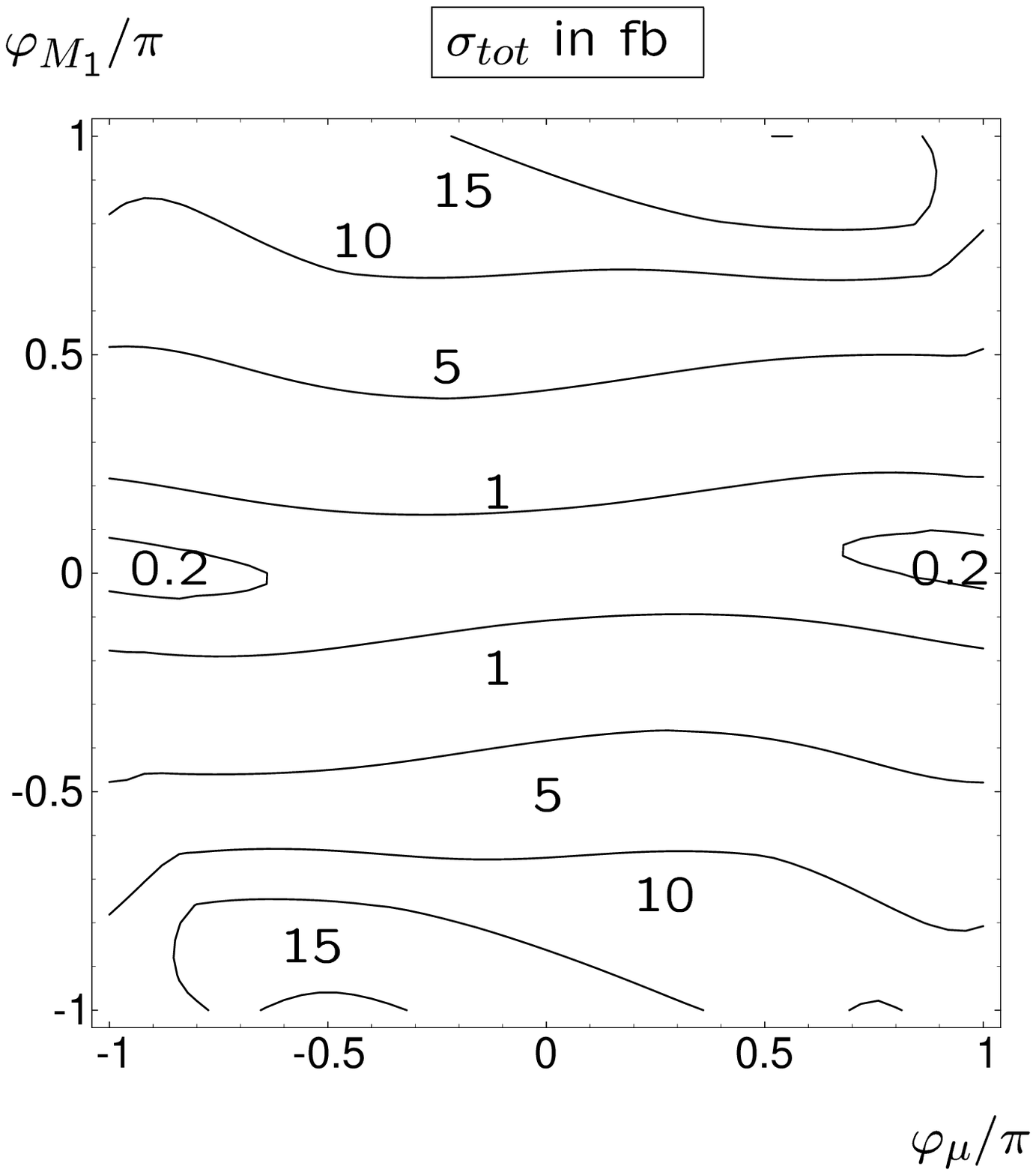}
\put(-180,0){(a)}
\end{minipage}\hspace{4pc}%
\begin{minipage}{16pc}
\includegraphics[height=16pc,width=16pc]{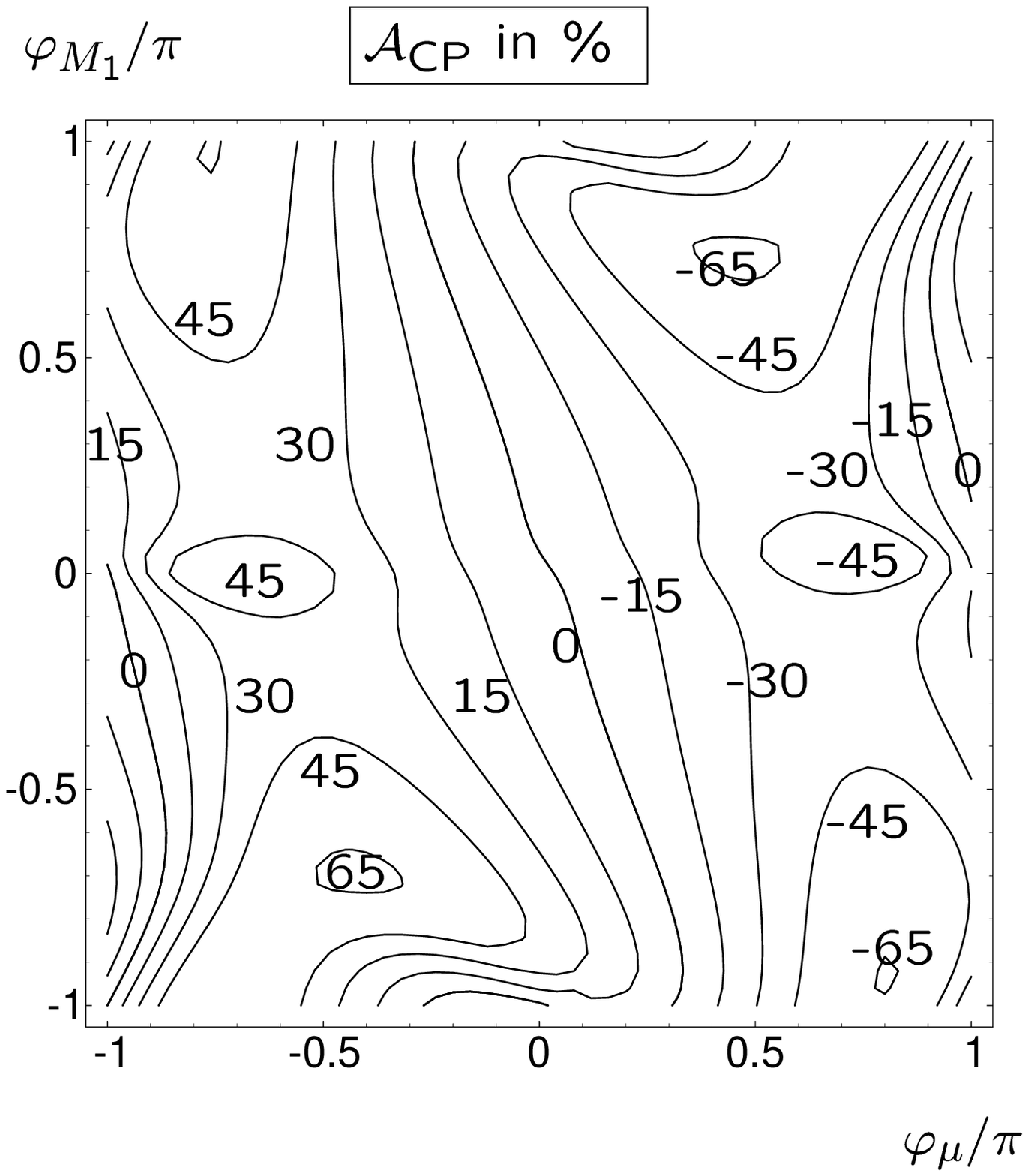}
\put(-180,0.){(b)}
\end{minipage} 
\caption{\label{FigTau} 
     Contour lines of (a) the cross section
     $\sigma = \sigma_P(e^+e^-\to \tilde\chi^0_1\tilde\chi^0_2)\times 
     {\rm BR}(\tilde\chi^0_2\to\tilde\tau_1^+\tau^-)$  
     and (b) the CP asymmetry ${\mathcal A}_{\rm CP}$
     in the $\varphi_{\mu}$--$\varphi_{M_1}$ plane, for
     $M_2=400$ GeV, $|\mu|=300$ GeV, 
    $\tan \beta=5$, $\varphi_{A_{\tau}}=0$, $A_{\tau}=250$~GeV, 
    at $\sqrt s = 500$~GeV, and polarized beams
    $(P_{e^-},P_{e^+})=(-0.8,0.6)$~\cite{Bartl:2003gr}. }
\end{figure}

As a next example, we show that final state particle polarizations
can give large asymmetries. The transverse polarization $\vec s_\tau$ 
of the tau in the neutralino decay  
\begin{eqnarray}
\tilde\chi^0_2 \to \tilde\tau_1^\pm + \tau^\mp
\label{TAUdecay}
\end{eqnarray}
can be used to form the triple product
${\mathcal T} = (\vec p_\tau \times \vec p_{e^-})
        \cdot \vec s_\tau$~\cite{Bartl:2003gr,Choi:2003pq}.
In Fig.~\ref{FigTau}, we show the phase dependence of the
corresponding asymmetry ${\mathcal A}_{\rm CP}$, see Eq.~(\ref{eq:CPasymmetry}), 
and the cross section. The asymmetry is very sensitive 
on the phases and reaches up to $65\%$.
Note, that although we have shown only the $\varphi_{\mu}$--$\varphi_{M_1}$
dependence of  ${\mathcal A}_{\rm CP}$, it is also very sensitive to the phase
$\varphi_{A_{\tau}}$ in the stau sector, in particular for 
$|A_{\tau}|\gg |\mu|\tan\beta$~\cite{Bartl:2003gr}.

\medskip

With these examples, we close the discussion of
triple products at the ILC. Note however, that there is a vast amount
of analyses that study triple products in two-body decays of neutralinos~\cite{NEUT2}
and charginos~\cite{CHAR2}, three-body decays of neutralinos~\cite{NEUT3} 
and charginos~\cite{CHAR3}, and also CP-odd observables with transversely polarized beams
at the ILC~\cite{Trans}. For a general discussion of the neutralino system with CP phases,
see Ref.~\cite{NEUT}.


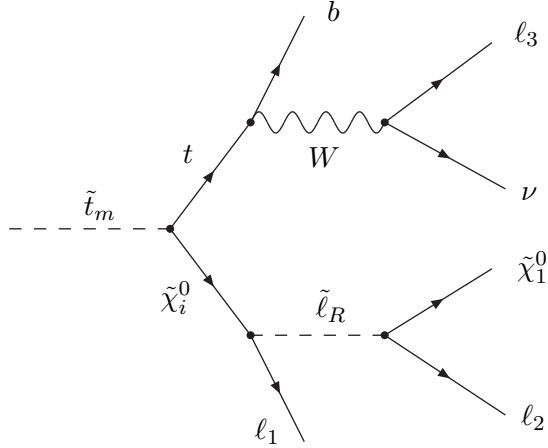
\begin{figure}[t]
                \scalebox{1}{
\setlength{\unitlength}{1cm}
\begin{picture}(10,9)(-3.4,0)
          \DashLine(-10,80)(50,80){5}
          \Vertex(50,80){1.5}
          \ArrowLine(50,80)(80,120)
          \Vertex(80,120){1.5}
          \ArrowLine(80,120)(100,160)
           \Photon(80,120)(130,120){4}{4}
           \Vertex(130,120){1.5}
            \ArrowLine(130,120)(170,150)
            \ArrowLine(130,120)(175,95)
           \ArrowLine(50,80)(80,40)
           \Vertex(80,40){1.5}
           \ArrowLine(80,40)(100,0)
           \DashLine(80,40)(130,40){5}
           \Vertex(130,40){1.5}
          \ArrowLine(130,40)(170,65)
          \ArrowLine(130,40)(175,10)
         \put( 0.5,3.0){ $\tilde t_m $}
         \put( 1.51,1.8){ $\tilde\chi_i^0 $}
         \put( 2.75,0.03){ $\ell_1$}
         \put(3.55,1.65){ $\tilde\ell_R$}
         \put(6.2,2.2){ $\tilde\chi_1^0$}
         \put(6.25,0.22){ $ \ell_2$}
         \put( 1.8,3.7){ $t$}
         \put(3.45,3.64){ $W$}
         \put( 3.7,5.58){ $b$}
         \put(6.25, 3.2){ $\nu $}
         \put(6.16,5.3){ $\ell_3$}
\end{picture}
}
\caption{Schematic picture of the two-body stop quark decay-chain.}
\label{Fig:decayStop}
\end{figure}

\begin{figure}[h]
\begin{minipage}{16pc}
\includegraphics[height=16pc, width=16pc]{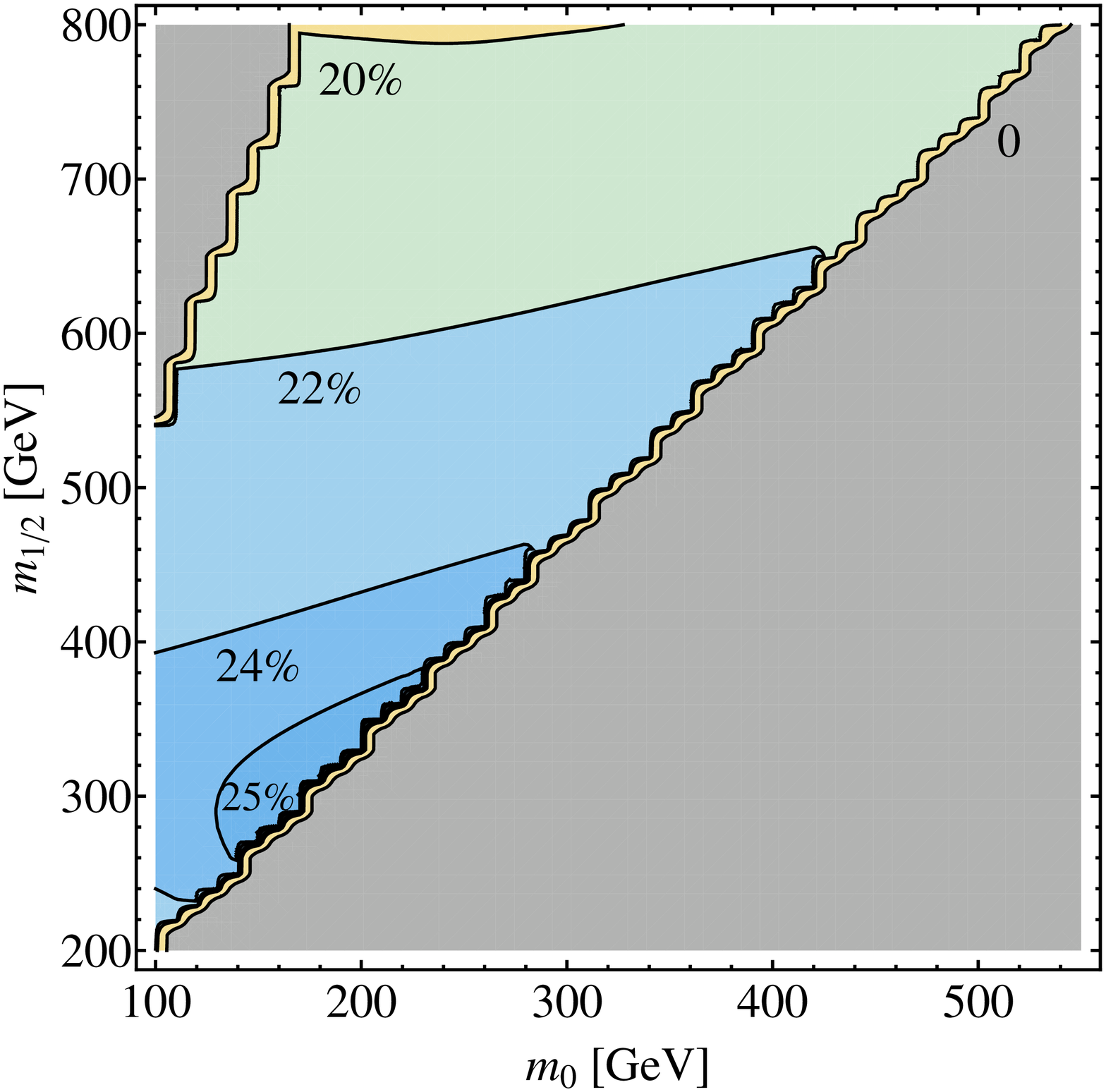}
\put(-80,40){$m_{\tilde\ell_R}> m_{\tilde\chi_2^0}$}
\put(-180,0){(a)}
\end{minipage}\hspace{3pc}%
\begin{minipage}{16pc}
\includegraphics[height=17pc,width=19pc]{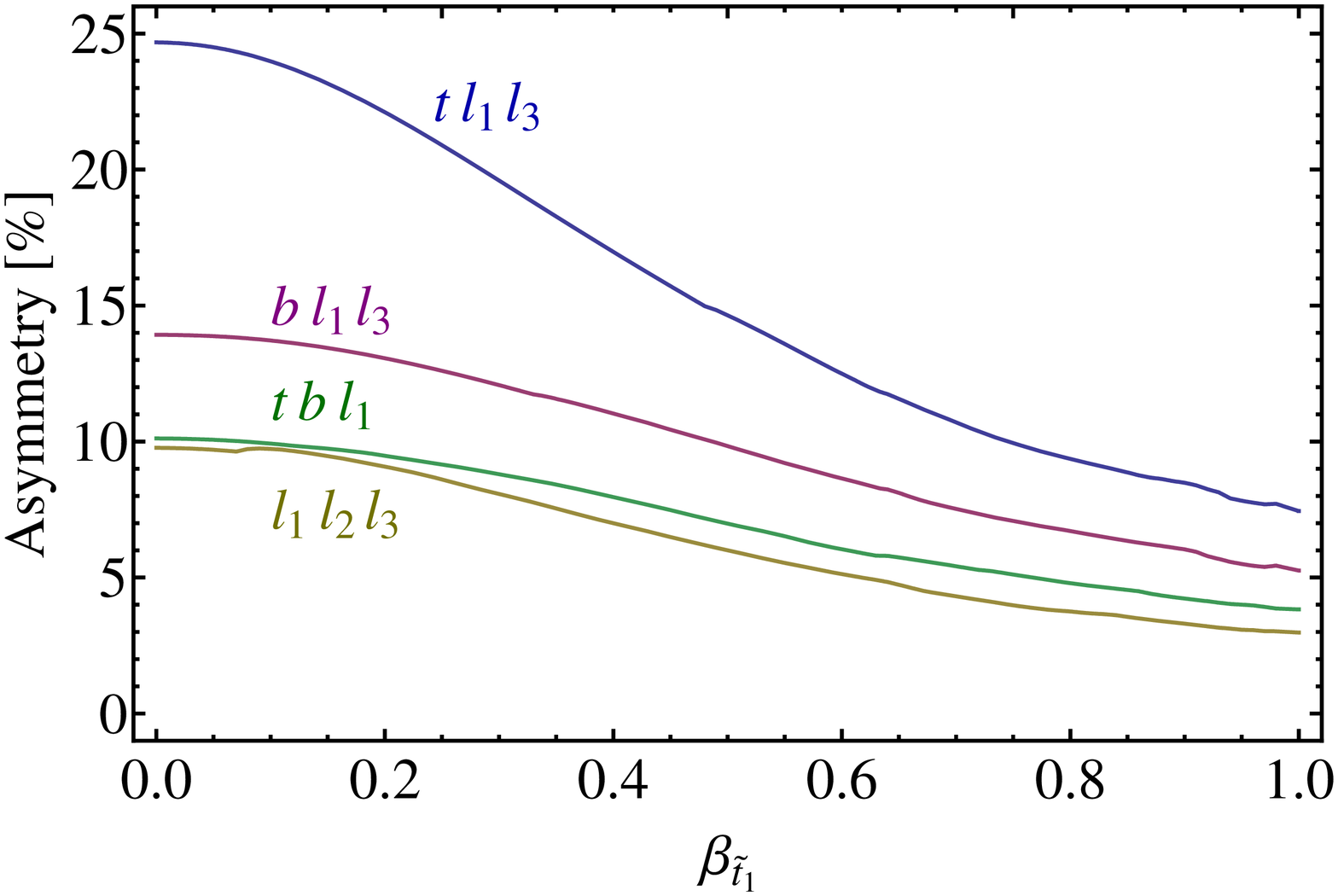}
\put(-180,0){(b)}
\end{minipage} 
\caption{\label{FigStop} 
        (a)~Contour lines in the $m_0$--$m_{1/2}$ plane of the 
         asymmetry ${\mathcal A}_{\rm T}$, Eq.~(\ref{eq:Tasymmetry}), 
         of the triple product 
        ${\mathcal T} = (\vec p_{t} \times \vec p_{\ell_1}) \cdot \vec p_{\ell_3}$
        for the $\tilde t_1\to t\tilde\chi_2^0$ decay chain as shown in Fig.~\ref{Fig:decayStop},
        in the stop rest-frame, i.e., for boost
        $\beta_{\tilde t_1} = |\vec p_{\tilde t_1}|/E_{\tilde t_1}=0$.
        The MSUGRA parameters are $\tan \beta=10$, $|A_{0}|=100 $ GeV,
        with added phases $\varphi_{A_0}= 0.64\pi$, $\varphi_{\mu}=\varphi_{M_1}=0$
        at the weak scale.
        The  upper left corner is excluded by    
        $m_{\tilde\chi_1^0}>m_{\tilde\ell_R }$.
        (b)~The boost 
        dependence of the various asymmetries for
        different combinations of the triple product momenta, 
        with MSUGRA parameters as in (a), and
        $m_0=100$ GeV, $m_{1/2}=300$~GeV~\cite{stop}.
        }
\end{figure}

\section{Numerical examples for the LHC}
\smallskip

We discuss the top squark decay at the LHC~\cite{Bartl:2004jr,stop,Ellis:2008hq}
\begin{eqnarray}
\tilde t_m \to t + \tilde\chi^0_i; \quad m=1,2; \quad i=2,3,4;
\label{eq:decayStop}
\end{eqnarray}
with the subsequent two-body decay chains of the neutralino
and the top as shown in Fig.~\ref{Fig:decayStop}.
Only the spin-spin correlations of the neutralino and the top
are sensitive to the imaginary part of the
product of the left and right $\tilde t_m$--$t$--$\tilde\chi^0_i$
couplings,
${\rm Im}\{ a^{\tilde t}_{mi} (b^{\tilde t}_{mi})^\ast \}$,
which depend on the phases
$\varphi_{A_t}$, $\varphi_{\mu}$ and $\varphi_{M_1}$~\cite{Bartl:2004jr}.
The final state particle momenta
$\vec p_{t}, \vec p_{b} ,\vec p_{\ell_1},\vec p_{\ell_2},\vec p_{\ell_3}$
can be used to define various triple products and their corresponding asymmetries.
In the following, we concentrate on the asymmetries in the two-body decays of 
the neutralino~\cite{Bartl:2004jr, stop}.
Triple product asymmetries in the three-body stop decay have been studied 
in Ref.~\cite{Bartl:2002hi}.

\clearpage

   Note that if the decay (or the transverse polarization) of the top is not 
   taken into account, the spin-spin correlations are lost. 
   Then only CP asymmetries can be obtained, which are sensitive to the
   CP phases $\varphi_{\mu}$ and $\varphi_{M_1}$ alone, which enter solely from the 
   neutralino decay.
   Still in that case, a three-body neutralino
   decay is required~\cite{Ellis:2008hq, sfermNEUT3}, 
   or a two body-decay chain with an intermediate $Z$-boson~\cite{Bartl:2003ck}. 
   However, the sfermion then merely serves as a production channel for neutralinos,
   to analyze their CP properties through their subsequent 
   decays~\cite{Ellis:2008hq, sfermNEUT3,Bartl:2003ck}.

\smallskip

In Fig.~\ref{FigStop}(a), we now show contourlines of the
asymmetry  ${\mathcal A}_{\rm T}$  from the triple product
 ${\mathcal T} = (\vec p_{t} \times \vec p_{\ell_1}) \cdot \vec p_{\ell_3}$
in the $m_0$--$m_{1/2}$ plane.
We choose an MSUGRA-inspired scenario, with the
input parameters $m_0$, $m_{1/2}$, $\tan \beta$, $A_{0}$, at the GUT
scale, to obtain the low energy parameters.
We then add the CP-violating phases  $\varphi_{A_0}$, $\varphi_{\mu}$, $\varphi_{M_1}$. 
We see that the asymmetry can reach up to $25\%$, values which have also
been found in Ref.~\cite{Bartl:2004jr}. 
However, the asymmetry has been evaluated in the rest-frame of the stops.
At the LHC the stops will be boosted, and the corresponding asymmetries 
will be reduced~\cite{Ellis:2008hq}.
In Fig.~\ref{FigStop}(b) we show their stop boost dependence. 
We also show other asymmetries, which are obtained using different combinations 
of momenta for the triple products. 

\smallskip

Finally we want to shortly comment on the tri-lepton signal at the LHC.
If a produced pair of a chargino and a neutralino both decay 
leptonically, three leptons and missing energy will form a 
distinctive signal with low QCD background~\cite{Baer:1994nr}.
A triple product of the three leptons has been analyzed at the Tevatron, 
however only small asymmetries have been obtained~\cite{trilepton}.
For the LHC, a systematic analysis of CP observables in the tri-lepton 
signal is planned~\cite{plan}.

\begin{figure}[t]
                \scalebox{1}{
\setlength{\unitlength}{1cm}
\begin{picture}(10,6)(-3.4,0)
           \ArrowLine(30,90)(0,120)
           \ArrowLine(0, 40)(30,70)
          \ArrowLine(50,90)(80,120)
          \ArrowLine(80, 40)(50,70)
           \CArc(40,80)(14.5,0,360)
         \Vertex(80,120){1.5}
          \ArrowLine(100,160)(80,120)
           \DashLine(80,120)(130,120){5}
           \Vertex(130,120){1.5}
            \ArrowLine(170,150)(130,120)
            \ArrowLine(130,120)(175,95)
           \Vertex(80,40){1.5}
           \ArrowLine(100,0)(80,40)
           \DashLine(80,40)(130,40){5}
           \Vertex(130,40){1.5}
          \ArrowLine(170,65)(130,40)
          \ArrowLine(130,40)(175,10)
         \put( 0.45,3.85){ $\bar d $}
         \put( 1.60,3.85){ $\tilde\chi_j^+$}
         \put( 0.45,1.6){ $  u $}
         \put( 1.60,1.6){ $\tilde\chi_i^0 $}
         \put( 2.65,0.03){ $\ell_1^+$}
         \put(3.55,1.65){ $\tilde\ell^-$}
         \put(6.2,2.2){ $\tilde\chi_1^0$}
         \put(6.25,0.03){ $ \ell_2^-$}
         \put(3.45,3.64){ $\tilde\nu_\ell$}
         \put( 3.7,5.58){ $\ell_3^+$}
         \put(6.25, 3.2){ $\nu $}
         \put(6.16,5.3){ $\tilde\chi_1^0$}
\end{picture}
}
\caption{Schematic picture of neutralino-chargino production and decay.
 }
\label{Fig:process}
\end{figure}
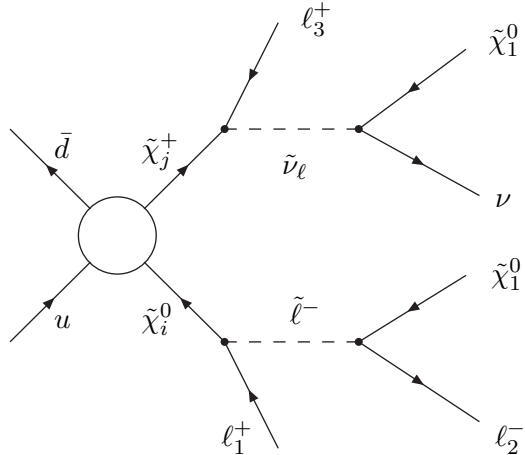

\section{Summary and conclusions}
\smallskip

We have shown how triple products of particle momenta or spins can be used
to define asymmetries, which are sensitive to the CP phases of the MSSM.
Such asymmetries can be large, since they are already present at the tree level.
For example, for neutralino production and decay at the ILC, and squark decays at the
LHC, we have shown that the asymmetries reach up to $60\%$ and $25\%$, respectively.
As the asymmetries can be sensitive to small phases of order $0.1\pi$, they will
be an ideal tool to measure or constrain SUSY CP phases at colliders, independently
from low energy measurements of electric dipole moments.
We hope that our theoretical studies motivate detailed  
experimental studies, taking into account backgrounds and event reconstruction 
efficiencies, to resolve the question whether SUSY CP phases can 
indeed be measured at the ILC and LHC.

 \ack

I would like to thank
A.~Bartl, S.~Bornhauser, F.~Deppisch,  M.~Drees, H.~Dreiner, O.~J.~P.~'Eboli, H.~Fraas, 
K.~Hohenwarter-Sodek, T.~Kernreiter, J.~S.~Kim, S.~Kulkarni,
W.~Majerotto, A.~Marold, F.~v.d.~Pahlen, and M.~Terwort
for valuable collaborations.
This work was supported by MICINN project FPA.2006-05294.

\end{document}